# In-line optical subtraction using a differential Faraday rotation spectrometer for $^{15}$NO/$^{14}$NO isotopic analysis

ERIC J. ZHANG,[1,2]* DANIEL M. SIGMAN,[3] GERARD WYSOCKI[1]

[1]*Department of Electrical Engineering, Princeton University, Princeton, NJ 08544, USA*
[2]*IBM Thomas J. Watson Research Center, Yorktown Heights, NY 10598, USA (current affiliation)*
[3]*Department of Geosciences, Princeton University, Princeton, NJ 08544, USA*

*Corresponding author: eric.jh.zhang@ibm.com*

**We present a dual-modulation Faraday rotation spectrometer with in-line optical subtraction for differential measurement of nitric oxide (NO) isotopologues. In-situ sample referencing is accomplished via differential dual-cell measurements, with 3.1 ppbv·Hz$^{-1/2}$ ($^{15}$NO) sensitivity through 15 cm optical path length. Our system operates at 1.9× the shot-noise limit, with a minimum fractional absorption of 1.8×10$^{-7}$ Hz$^{-1/2}$. Differential measurement of both $^{14}$NO and $^{15}$NO are shown, yielding ~20 dB magneto-optical suppression. Noise analysis demonstrates stability of the differential signal up to ~500 s, with normalized ratiometric precision of 3.0 ‰·Hz$^{-1/2}$ using 1 ppmv $^{15}$NO (or 272 ppmv $^{14}$NO at natural abundance). We rigorously model our differential method and demonstrate the utility of in-line calibration for precise isotopic ratiometry.**

Precision isotopic analysis of nitric oxide ($^{15}$NO, $^{14}$NO) and of related nitrogen forms is an emerging technique for medical and environmental diagnostics [1-4], as various biological and geochemical pathways exhibit isotopic fractionation [2, 4]. In biochemical applications, isotopic analysis may be used to identify signaling pathways [3] and pathologies via human metabolic studies [4], while environmental applications include analysis of nitrogen isotopes to determine anthropogenic contributions in global nitrogen cycling dynamics [1, 5, 6]. State-of-art techniques rely on isotope-ratio mass-spectrometry (IRMS) [7], thus limiting their use due to significant instrumentation and cost overhead. Sub-permil (‰) level precision is required to assess isotopic variations in environmental samples in order to investigate many of the processes of interest, be they natural [8] or anthropogenic [9]. Therefore, the demands for precision and accuracy are high, necessitating, for example, frequent referencing to well-defined isotopic standards. The challenge is to provide precise and calibrated isotopic analysis without the cost and instrumentation overhead of IRMS systems. NO is a particularly valuable target for the development of isotopic methods as many other important nitrogen forms can be converted into it [10].

Laser spectroscopy has garnered broad interest for the detection of isotopologues due to measurable shifts in rovibrational energies arising from atomic mass variations [11, 12]. However, spectral contamination due to interfering molecular species (e.g. $H_2O$ and volatile organic compounds) imposes strict sample purity requirements for accurate quantification. In cases where the analyte exhibits paramagnetism (as is the case for detection of a variety of free radicals common in chemical processes), Faraday rotation spectroscopy (FRS) has been utilized to ensure immunity against non-paramagnetic species that reside within its spectral vicinity [13-17]. This combination of spectral and paramagnetic selectivity ensures highly accurate species discrimination, and FRS has found increasing utilization in the measurement of chemical radicals in applications ranging from breath analysis [14] to environmental detection [15, 16] and even studies of fuel oxidation rates in combustion diagnostics [17]. At present, state-of-art FRS systems are able to achieve a minimum detectable polarization rotation ("noise equivalent angle") of $\theta_{NEA} \sim 10^{-8}$ rad·Hz$^{-1/2}$, which corresponds to performance typically in the range of 1~2× the fundamental shot-noise limit.

A variety of FRS techniques have been developed to reach ultrasensitive limits required for free-radical detection [13, 15, 18]. Dual-modulation (DM-FRS) of both magnetic field and laser frequency has proven effective in mitigating both spectral contamination (e.g. optical etalons) and low-frequency noise [14, 16]. Previously, DM-FRS was utilized for time-multiplexed isotopic studies of NO, yielding sub-permil precision [13, 16, 19]; however, ratiometric accuracy remains challenging and is predicated on the assumption that reference and sample measurements are acquired *identically*. Ideally, in-situ calibration should be performed to ensure that sensor drift does not impact the ratiometric result over time. Here, we present a differential DM-FRS (dDM-FRS) sensor with in-line optical polarization subtraction of sample and reference signals by reversing the sense of Faraday rotation. The in-situ calibration process avoids ratiometric offsets associated with imperfect calibration. Effective magneto-optical suppression is necessary for accurate in-line referencing, which yields 19.2 dB common-mode rejection ratio (CMRR) for $^{15}$NO and ~500 s of zero-drift stability at 1.9× the shot-noise limit. The isotopic ratio is determined by measuring only the sample/reference differential signal, yielding a ratiometric precision of 3.0 ‰·Hz$^{-1/2}$ (normalized to 1 ppmv $^{15}$NO, or 272 ppmv $^{14}$NO at natural abundance) which may be improved with signal averaging.

The dDM-FRS system is depicted in Fig. 1, following initial developments outlined in [13, 19]. A distributed-feedback quantum cascade laser (Alpes Lasers) is employed to target the $^{15}$N$^{16}$O Q(3/2) and $^{14}$N$^{16}$O P(19/2)e transitions at 1842.76 cm$^{-1}$ and 1842.95 cm$^{-1}$ respectively. Laser modulation is performed at $f_L$ = 50 kHz for mitigation of low-frequency noise. The solenoid is driven sinusoidally at $f_M$ = 100 Hz (Behringer EPQ900) with 169 Gauss amplitude for near-optimal Zeeman splitting of the $\Delta M_j = \pm 1$ states of the $^{15}$N$^{16}$O Q(3/2) transition at an operating pressure of 80 Torr. As the beam passes through the gases with Zeeman-split transitions, relative circular birefringence is induced between the right- and left-hand circular polarizations, thereby inducing a rotation in the linearly polarized beam. Polarization measurement occurs in a polarizer/analyzer configuration (P1 and P2 in Fig. 1(a)), where the two anti-reflection wire-grid polarizers (ISP Optics, POL-3-5-SI-25) are positioned across the sample and reference cells. Each cell is 15 cm with angled wedged-glass windows to prevent spurious back-reflections and spectral fringing. Both sample and reference cells are housed within the same solenoid to minimize magnetic field dissimilarities. The resulting optical signal is measured on a TEC-cooled MCT photodiode (Vigo PVI-3TE) and undergoes two-stage demodulation for the dual sidebands at $2 \times f_L \pm f_M$ via phase-sensitive detection (Zurich Instruments HF2LI). Line-locked measurements are performed using the partial reflection from P1, which is sent through a high-concentration cell (2.5 vol.% $^{14}$NO and 1.5 vol.% $^{15}$NO), and the resulting $3f_L$ WMS signal is utilized to selectively lock the DFB-QCL to the $^{14}$NO or $^{15}$NO transition. Optimization of the analyzer (P2) angle is accomplished by approximately equalizing the signal relative intensity noise ($\sigma_{RIN} = 2.89 \times 10^{-7}$ Hz$^{-1/2}$) and detector noise-equivalent power ($\sigma_{NEP} = 1.15 \times 10^{-12}$ W·Hz$^{-1/2}$), yielding an empirical optimized uncrossing angle of $\theta_{opt} \simeq 2.5°$. For an incident power (on the initial polarizer) of $P_0 = 2.0$ mW, this corresponds to 3.8 µW on the signal photodiode.

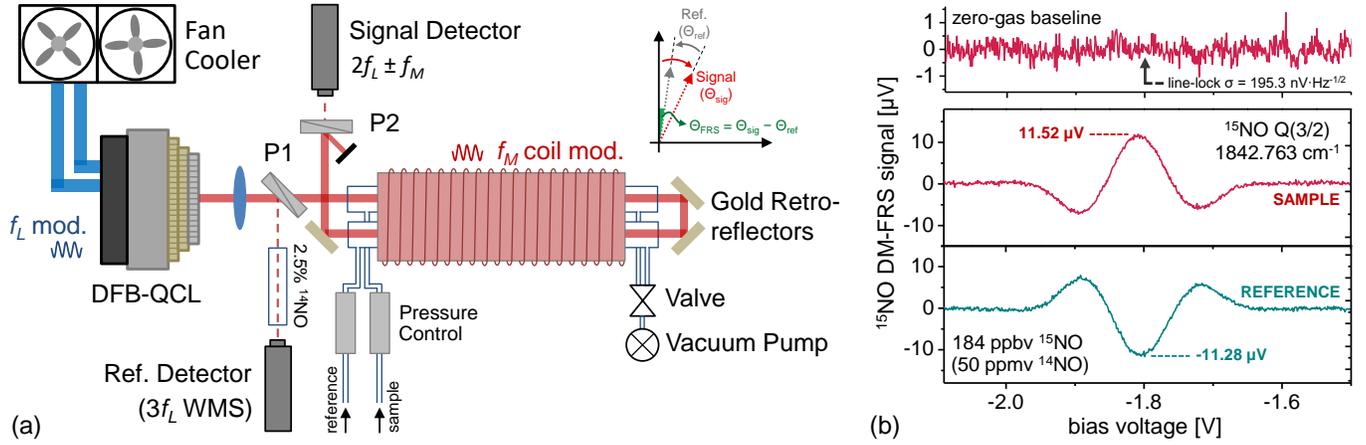

Fig. 1: (a) Layout of the dDM-FRS setup for polarization subtraction of sample and reference signals. The polarized beam (after P1) passes sequentially through sample and reference gas cells (80 Torr), and the retroreflector maintains the circular polarization sense of the beam. The opposite magnetic field (with respect to beam direction) in the reference cell induces reversal of Faraday rotation polarity, yielding a differential signal which is analyzed (P2) on the signal detector (schematically illustrated in the top right inset). (b) Spectral acquisition of the $^{15}$NO Q(3/2) transition through the sample cell (red spectrum), using a calibrated 184 ppbv $^{15}$NO cylinder. The green curve shows the corresponding reference cell spectrum, which is inverted due to the reversal of Faraday rotation polarity. The top panel shows ambient laboratory air measurement, used as zero-gas at natural abundance and demonstrates the near zero-baseline FRS spectrum. Noise analysis indicates an SNR of 59.0 Hz$^{-1/2}$, yielding a $^{15}$NO detection limit of 3.1 ppbv·Hz$^{-1/2}$.

As shown in the top right inset of Fig. 1(a), optical subtraction of the signal and reference gases is accomplished by use of a retroreflector consisting of two bare-gold mirrors aligned at a relative angle of 90°. The reflection off each gold mirror causes the reversal of the 'handedness' of circular polarization in the beam, and the combination of two mirrors ensures the polarization sense between sample and reference cells are identical. However, the opposite magnetic field in the reference (with respect to beam direction) causes the Faraday rotation to be reversed in polarity. This yields relative concentration between the signal and reference gas cells, i.e. in-line polarization subtraction which enables in-situ calibration without requiring separate reference measurements.

Fig. 1(b) demonstrates the effect of Zeeman-splitting polarity reversal on the measured Q(3/2) $^{15}$NO spectrum using gas from a certified 50 ppmv $^{14}$NO cylinder (N$_2$ balance), which provides 184 ppbv $^{15}$NO at natural abundance (0.367 vol.%). The sample (red) and reference (green) spectra were acquired by filling only one respective cell, yielding spectra of opposite polarity after signal demodulation. By filling both sample and reference simultaneously, signal subtraction may thus be performed (demonstrated in Fig. 2). The top panel in Fig. 1(b) shows the same measured spectrum for ambient air, which is effectively zero-gas at natural $^{15}$NO abundance. The absence of contaminating spectral features indicates the baseline-free nature of DM-FRS. Noise performance is assessed by line-locking to the $3f_L$ WMS zero-crossing of the high-concentration wavelength reference cell, yielding 195.3 nV·Hz$^{-1/2}$ Gaussian-noise limited performance [13]. Based on the peak signal amplitude of $V_{sig} = 11.52$ µV in Fig. 1(b), we calculate a SNR = 59.0 Hz$^{-1/2}$ or $^{15}$NO detection limit of 3.1 ppbv·Hz$^{-1/2}$, corresponding to a minimum fractional absorption $(\alpha L)_{min} = 1.83 \times 10^{-7}$ Hz$^{-1/2}$ (see Supplement C). The minimum detectable polarization rotation (i.e. noise-equivalent angle) is $\Theta_{NEA} = 1.70 \times 10^{-8}$ rad·Hz$^{-1/2}$, demonstrating performance comparable to state-of-art FRS systems [13, 15, 16]. Similar measurements for optimized $^{14}$NO DM-FRS signals yields 20.3 ppbv·Hz$^{-1/2}$ sensitivity [13]. Nominally, magnetic field optimization is performed to maximize the $^{15}$NO signal, resulting in a relative decrease in $^{14}$NO isotopic precision due to the sub-optimal Zeeman-split $^{14}$N$^{16}$O P(19/2)e transition. Nevertheless, natural abundance will yield ~270× greater concentrations of $^{14}$NO as compared with $^{15}$NO and thus the lower sensitivity to the major isotope negligibly impacts the precision of ratiometric analysis.

To demonstrate real-time optical polarization subtraction, a multi-stage line-locked measurement is depicted in Fig. 2(a), whereby the sample and reference gas cells are filled individually and then concurrently during successive dDM-FRS acquisition sequences. Between ~200 to 600 s, only sample gas is measured (184 ppbv $^{15}$NO), yielding a positive dDM-FRS signal with amplitude $[^{15}\tilde{N}]_S = 11.61$ µV (averaged over the step duration), and is similar to Fig. 1(b). Between ~850 to 1150 s, the identical gas is flown through only the reference cell, yielding a measured amplitude of $[^{15}\tilde{N}]_R = -11.21$ µV. The negative sign is measured directly after demodulation and is indicative of the opposite Zeeman split polarity of the $\Delta M_j = \pm 1$ states. The difference between sample and reference signal magnitude is primarily attributable to path length variations of the customized gas cells, which is within the glassware fabrication error. Between ~1150 to 1400 s, both sample and reference cells are filled simultaneously, resulting in the optical difference $^{15}\tilde{\eta} = 0.43$ µV. A null measurement is performed from 1500 s onwards, yielding $V_0 = 95$ nV and is used as the zero-gas baseline of our system. This minor non-zero deviation is a consequence of electromagnetic interference (EMI) from the solenoid, which may be reduced by improved high magnetic permeability shielding of the laser and cabling in future design iterations. To ascertain the magneto-optical common mode rejection ratio (CMRR), we account for this null offset in each of the measured signals, i.e. $\{^{15}\tilde{\eta}, [^{15}\tilde{N}]_S, [^{15}\tilde{N}]_S\} \to \{^{15}\tilde{\eta} - V_0, [^{15}\tilde{N}]_S - V_0, [^{15}\tilde{N}]_S - V_0\}$ which yields a $^{15}$NO suppression factor:

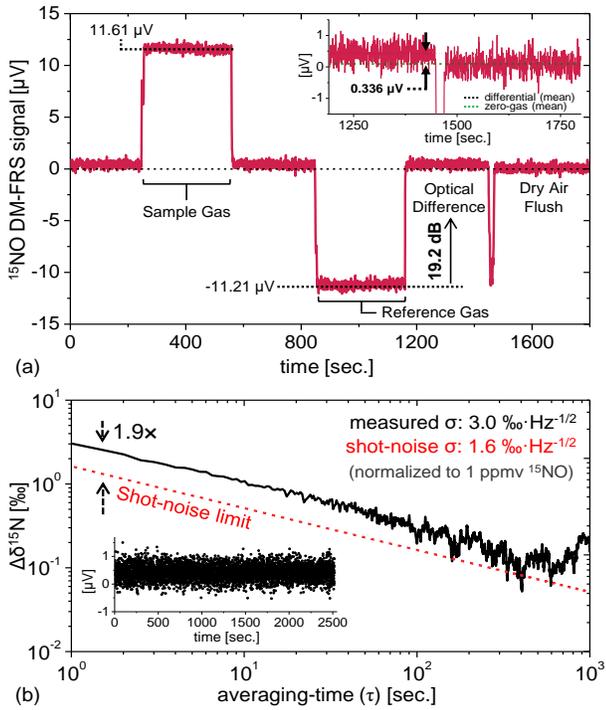

Fig. 2: (a) Real-time in-line differential $^{15}$NO line-locked signals. The step measurements involve injection of 184 ppbv $^{15}$NO into the 15 cm sample gas cell (11.61 µV), reference cell (-11.21 µV), and both sample and reference cells simultaneously. The inset showing a magnified plot of the differential measurement. Compensating for differences in sample/reference optical path lengths yields $^{15}$CMRR of -19.2 dB. (b) Allan deviation analysis of the $^{15}$NO differential signal in ratiometric permil ($\delta^{15}$N) units, which yields 3.0 ‰·Hz$^{-1/2}$ (at 1 ppmv $^{15}$NO) with white-noise performance up to ~500 s at 1.9× the shot-noise limit.

$$^{15}R_{CMRR} = \left| \frac{\langle ^{15}\tilde{\eta}(t) \rangle}{\langle [^{15}\tilde{N}]_R(t) \rangle} \right| = 1.2 \times 10^{-2} \quad (1)$$

Corresponding to $^{14}CMRR = 10 \cdot \log(^{15}R_{CMRR}) = -19.2$ dB. Similar analysis of the major isotope yields $^{14}R_{CMRR} = 3.4 \times 10^{-3}$ and $^{14}CMRR = -24.7$ dB. As described above, this finite CMRR is primarily attributable to path length differences between the sample and reference cells (and to lesser extent, magnetic field inhomogeneities due to unequal positioning with respect to the solenoid axis). Long-term stability of a $^{15}\tilde{\eta}_{meas}$ differential measurement is shown in Fig. 2(b), where the $^{15}$NO line-locked dDM-FRS signal is acquired over 2500 s (inset). Allan deviation analysis indicates noise performance at 1.9× the fundamental shot-noise limit, with white-noise limited performance up to ~500 s. The noise measurement is depicted in permil precision values (calculated using the dominant $\Omega_{O(1)}(t)$ contribution to $\delta^{15}$N in Eq. 3), yielding a precision of 3.0 ‰·Hz$^{-1/2}$, normalized to 1 ppmv $^{15}$NO, or 272 ppmv $^{14}$NO at natural abundance (0.367 vol. %). Given our sensor stability time of ~500 s, precisions approaching 0.1‰ are achievable with averaging and > 1 ppmv $^{15}$NO (> 272 ppmv $^{14}$NO) sample concentration in our dDM-FRS sensor, with linear increase in precision for increasing sample size. We note here that the impact of finite $^{15}R_{CMRR}$ is to contribute an additive offset to $\delta^{15}$N (derived in Eq. (3)), which is an easily correctable offset and remains stable up to the zero-drift time of the Allan deviation plot shown in Fig. 2(b).

In prior work [16], one of the principal challenges in achieving precise and accurate isotopic ratiometry involved the temporally separate and therefore non-identical acquisition conditions of sample and reference gases. Therefore, even minor variations in flow fractionation offsets easily dominate the overall ratiometric errors. In contrast, our dDM-FRS system is uniquely capable of simultaneous reference and sample difference measurements, and therefore in-situ, real-time calibration. In what follows, our goal is to rigorously demonstrate the benefit of in-line referencing and we derive an expression for the isotopic ratio based on sample/reference differential signals $^{14}\tilde{\eta}$ and $^{15}\tilde{\eta}$. We begin from the definition:

$$\delta^{15}N(t) = \left( \frac{\kappa_{15}}{\kappa_{14}} \cdot \frac{[^{15}\tilde{N}]_S(t)/[^{14}\tilde{N}]_S(t)}{[^{15}\tilde{N}]_R(t)/[^{14}\tilde{N}]_R(t)} - 1 \right) \times 10^3 \text{ ‰} \quad (2)$$

where $\kappa_{14}$ and $\kappa_{15}$ correspond to relative sample/reference detection sensitivities for $^{14}$NO and $^{15}$NO respectively and are ideally unity when sample and reference conditions are identical. We assume that non-zero offsets ($V_0$) are accounted for in all measured signals. Under the assumption that $^{14}\tilde{\eta} \ll [^{14}\tilde{N}]_R$ (i.e. sample and reference $^{14}$NO are similar), we arrive at an equivalent expression (for derivations, see Supplement A):

$$\delta^{15}N(t) = \frac{1+^{14}R_{CMRR}}{1+^{15}R_{CMRR}} \cdot \left\{ \begin{array}{l} \Omega_{O(1)} - \Omega_{O(2)} - \Omega_{O(3)} \\ +\tau(^{14}R_{CMRR}, ^{15}R_{CMRR}) \end{array} \right\} \times 10^3 \text{ ‰} \quad (3)$$

where we have defined (see Supplement B):

$$^{15}R_{CMRR} \equiv \frac{1-\kappa_{15}}{\kappa_{15}} \quad , \quad ^{14}R_{CMRR} \equiv \frac{1-\kappa_{14}}{\kappa_{14}} \quad (4)$$

which can be shown to be equivalent to the operational definition given in Eq. (1). $\Omega_{O(1)}(t), \Omega_{O(2)}(t), \Omega_{O(3)}(t)$ denote terms of increasing order (decreasing significance) to $\delta^{15}$N:

$$\Omega_{O(1)} = \frac{^{15}\tilde{\eta}(t)}{[^{15}\tilde{N}]_R(t)} \; , \; \Omega_{O(2)} = \frac{^{14}\tilde{\eta}(t)}{[^{14}\tilde{N}]_R(t)} \; , \; \Omega_{O(3)} = \frac{^{15}\tilde{\eta}(t)}{[^{15}\tilde{N}]_R(t)} \cdot \frac{^{14}\tilde{\eta}(t)}{[^{14}\tilde{N}]_R(t)} \quad (5)$$

The final term $\tau(^{14}R_{CMRR},^{15}R_{CMRR})$ in Eq. (3) accounts for finite CMRR, which manifests as an additive offset to $\delta^{15}$N, and can be quantified through methods depicted in Fig. 2(a):

$$\tau(^{14}R_{CMRR}, ^{15}R_{CMRR}) = \frac{^{14}R_{CMRR} - ^{15}R_{CMRR}}{1+^{14}R_{CMRR}} \quad (6)$$

Nominally, only the first-order error term contributes to the imprecision in Eq. (3), which yields (see Supplement D) [13]:

$$\Delta[\delta^{15}N(t)] = \frac{1+^{14}R_{CMRR}}{1+^{15}R_{CMRR}} \cdot \frac{\Lambda_\sigma \cdot \sigma_{shot}}{^{15}\tilde{\eta}(t)} \cdot \Omega_{O(1)} \cdot 10^3 \text{ ‰} \quad (7)$$

where we define $\Lambda_\sigma = 1.9$ as the performance factor above the fundamental shot-noise limit as shown in Fig. 2(b). Eq. (7) yields a precision of $\Delta[\delta^{15}N(t)] \simeq 3.1$ ‰·Hz$^{-1/2}$ (at 1 ppmv $^{15}$NO using $\sigma_{shot} = 102.7$ nV·Hz$^{-1/2}$) and is consistent with the value determined via Allan deviation analysis. Eqs. (3) and (7) indicate two key features of our dDM-FRS measurement. First, assuming we control $^{14}\tilde{\eta} \ll [^{14}\tilde{N}]_R$, and that $[^{14}\tilde{N}]_R$, $[^{15}\tilde{N}]_R$ and CMRR are known precisely (see Supplement D and E), $\delta^{15}$N depends only on $^{14}\tilde{\eta}(t)$ and $^{15}\tilde{\eta}(t)$ (Eqs. (3-6)), which was the initial goal in our differential formulation. In other words, our differential signal provides all information required to determine $\delta^{15}$N in real-time. Second, measurement uncertainties are dominated by $^{15}\tilde{\eta}(t)$ and contribute only *fractional* errors to $\delta^{15}$N (Eq. (7)). We contrast this with temporally separate reference and sample measurements [16], where even minor variations in the reference ratio (e.g. small fractionation offsets in the gas flow system) may cause intolerable offsets in ratiometric analysis.

In the formulation of Eq. (3), we have assumed two caveats: (i) the CMRR remains stable over the measurement duration and (ii) $^{14}\tilde{\eta} \ll [^{14}\tilde{N}]_R$. In (i), the CMRR offset term $\tau$ (Eq. (6)) is identical to the ~500 s stability time derived from Allan-deviation analysis of $\Omega_{O(1)}$ (Fig. 2(b)); thus no recalibration is required within the zero-drift time. In (ii) the reference and sample gases must be similar in concentration. Empirically, real-time dilution tracking may be used to ensure $^{14}\tilde{\eta}$ remains adequately small and will be implemented in future system prototypes.

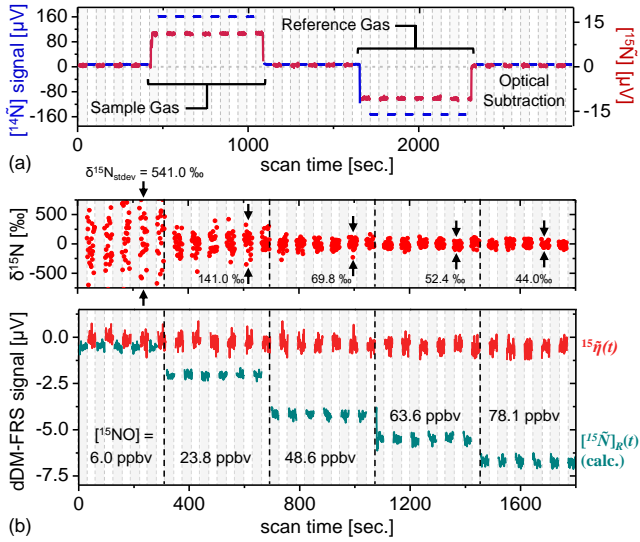

Fig. 3: (a) Line-switched measurements (i.e. sequential line-locking) alternating between $^{15}$NO (red) and $^{14}$NO (blue) at 100 s intervals, showing sample, reference, and differential signals. (b) A five-step dilution measurement, with differential signal $^{15}\tilde{\eta}(t)$ (red) compared against calculated $[^{15}\tilde{N}]_R(t)$ (cyan). Ratiometric $\delta^{15}$N is calculated via Eq. (3), with improved precision for increasing $^{15}$NO concentrations up to 78.1 ppbv, yielding a normalized precision of 3.4 ‰·Hz$^{-1/2}$ (at 1 ppmv $^{15}$NO) consistent with the value determined from noise analysis in Fig. 2(b).

Fig. 3(a) shows a measurement of both $^{14}$NO and $^{15}$NO isotopes on our dDM-FRS system with a single laser, by sequentially line-locking to the $^{15}$N$^{16}$O Q(3/2) and $^{14}$N$^{16}$O P(19/2)e transitions [16, 19], using the 3$f_L$ WMS reference photodiode signal (Fig. 1(a)). An example is shown in Fig. 3(a), which repeats Fig. 2(a) with additional line-switching at 100 s intervals, allowing measurement of time-multiplexed $^{14}$NO and $^{15}$NO signals. The line-lock time-constant is ~5 s, yielding a measurement duty cycle of 47.5 % for each isotope.

To demonstrate the ratiometric precision dependence on varying NO concentration, a five-step dilution experiment is performed in Fig. 3(b), with the diluted gas flowing into both sample and reference cells. Line-switching is performed at a faster rate of 30 s intervals for higher frequency of quasi-simultaneous signal acquisition from both isotopologues. Using the differential $^{14}$NO signal $^{14}\tilde{\eta}(t)$, we calculate $[^{14}\tilde{N}]_R(t)$ from our known $^{14}R_{CMRR}$ and reference gas concentration, which is used to calculate $[^{15}\tilde{N}]_R$ (cyan points) based on natural abundance. The differential $^{15}\tilde{\eta}(t)$ is measured directly (red points) and is effectively null due to the magneto-optical suppression. The top panel shows calculated $\delta^{15}$N, demonstrating progressive improvement of ratiometric precision as $^{15}$NO concentration is increased. As an experimental verification of our derived model, all terms ($\Omega_{O(1)}(t)$, $\Omega_{O(2)}(t)$, $\Omega_{O(3)}(t)$ and $\tau(^{14}R_{CMRR},^{15}R_{CMRR})$) in Eq. (3) have been utilized in the determination of $\delta^{15}$N. At 78.1 ppbv $^{15}$NO, we measure a ratiometric precision of 44‰, corresponding to a concentration-normalized precision of 3.4 ‰·Hz$^{-1/2}$ (at 1 ppmv $^{15}$NO) in good agreement with Allan deviation analysis in Fig. 2(b). In our present configuration, sample concentrations > 3.4 ppmv $^{15}$NO will thus enable sub-permil precisions for 1 s averaging time. In cases where averaging up to the instrument stability time (~500 s) is permitted, this concentration constraint is relaxed to merely $^{15}$NO ~ 50 ppbv ($^{14}$NO ~ 14 ppmv). Finally, we note that $\delta^{15}$N in Fig. 3(b) remains null, independent of NO concentration, indicating that our dDM-FRS system does not introduce fractionation artefacts and may therefore be effectively utilized for high-precision applications.

In our first-generation dDM-FRS prototype, we demonstrate in-situ calibrated isotopic ratiometry of NO using sample and reference differential signals. Noise analysis shows performance at 1.9× the shot-noise limit (~500 s zero-drift time), with $\Theta_{NEA} = 1.70 \times 10^{-8}$ rad·Hz$^{-1/2}$ and minimum fractional absorption $(\alpha L)_{min} = 1.83 \times 10^{-7}$ Hz$^{-1/2}$, on-par with state-of-art FRS systems. In-line magneto-optical subtraction (CMRR ~20 dB) is used for in-situ calibration, with 3.0 ‰·Hz$^{-1/2}$ ratiometric precision (normalized to 1 ppmv $^{15}$NO, or 272 ppmv $^{14}$NO). Our results show the dDM-FRS system to be free of fractionation artefacts. Isotopic ratiometry from dilution step measurements validates our in-line calibration model. Future work will focus implementation of active dilution tracking for real-time ratiometry, in addition to implementing a second laser to simultaneously measure both isotopologues and mitigate the impact of measurement duty cycling on averaging precision.

**Funding.** The authors acknowledge funding support from the Eric and Wendy Schmidt Transformative Technology Fund, the Walbridge Fund (Princeton Environmental Institute), and the Natural Sciences and Engineering Council of Canada.

**Acknowledgments.** The authors thank Q. Ji and F. Nuruzzaman for helpful discussions, and M. J. Souza for custom glassware construction.